\documentclass{ws-ijmpcs}
\usepackage{amssymb,graphicx}
\usepackage{epstopdf}
\DeclareGraphicsExtensions{.pdf,.eps,.png,.jpg,.mps}
\begin{document}
\markboth{M. Januar, A. Sulaiman and L.T Handoko}
{Nonlinear conformation of secondary protein folding}

%
\catchline{}{}{}{}{}
%

\def\grad{\mathbf{\nabla}}
\def\om{\mathbf{\omega}}
\def\v{\mathbf{v}}
\def\x{\mathbf{x}}
\def\u{\mathbf{U}}
\def\f{\mathbf{F}}
\def\s{\mathbf{S}}
\def\en{\mathbf{E}}
\def\r{\mathbf{r}}
\def\mb{\bar{m}}
\def\mt{\tilde{m}}
\def\lt{\tilde{\lambda}}
\def\xp{{x^\prime}}
\def\l{{\cal L}}
\def\cd{{\cal D}}
\def\h{{\cal H}}
\def\f{{\cal F}}
\def\z{{\cal Z}}
\def\s{{\cal S}}
\def\e{{\cal E}}
\def\p{{\cal P}}
\def\cd{{\cal D}}
\def\cj{{\cal J}}
\def\pd{\partial}
\def\d{\mathrm{d}}
\def\j{\mathbf{J}}
\def\jt{\tilde{J}}
\def\lt{\tilde{\lambda}}
\def\mt{\tilde{m}}
\def\at{\tilde{\alpha}}
\def\jtv{\mathbf{\tilde{J}}}
\def\exp{\mathrm{exp}}
\def\ex{\mathrm{e}}
\def\be{\begin{equation}}
\def\ee{\end{equation}}
\def\bea{\begin{eqnarray}}
\def\eea{\end{eqnarray}}
\def\ie{\textit{i.e.} }
\def\etal{\textit{et.al.} }

\title{\bf NONLINEAR CONFORMATION OF SECONDARY PROTEIN FOLDING}

\author{M. JANUAR}
\address{Department of Physics, University of Indonesia,
Kampus UI Depok, Depok 16424, Indonesia}

\author{A. SULAIMAN}
\address{Badan Pengkajian dan Penerapan Teknologi, BPPT Bld. II (19$^{\rm th}$
floor), Jl. M.H. Thamrin 8, Jakarta 10340, Indonesia\\
asulaiman@webmail.bppt.go.id, sulaiman@teori.fisika.lipi.go.id}

\author{L.T. HANDOKO}
\address{Group for Theoretical and Computational Physics,
Research Center for Physics, Indonesian Institute of Sciences,
Kompleks Puspiptek Serpong, Tangerang, Indonesia\\
handoko@teori.fisika.lipi.go.id, handoko@fisika.ui.ac.id,
laksana.tri.handoko@lipi.go.id}
\address{Department of Physics, University of Indonesia,
Kampus UI Depok, Depok 16424, Indonesia}

\maketitle

\begin{history}
\received{Day Month Year}
\revised{Day Month Year}
\end{history}

\begin{abstract}
A model to describe the mechanism of conformational dynamics in secondary
protein based on matter interactions is proposed. The approach deploys
the lagrangian method by imposing certain symmetry breaking. The protein 
backbone is initially assumed to be nonlinear and represented by the 
Sine-Gordon equation, while the nonlinear external bosonic sources is
represented by $\phi^4$ interaction. It is argued that the nonlinear source
induces the  folding pathway in a different way than the previous work with
initially linear backbone. Also, the nonlinearity of protein backbone decreases
the folding speed.

\end{abstract}

\ccode{PACS numbers: 87.15.ad, 87.15.Cc, 87.15.hm}

\section{Introduction}

It is known that the nonlinear excitations play an important role
conformational dynamics. For instance, the effective bending rigidity of a
biopolymer chain could lead to a buckling instability \cite{mingaleev}. Some
models have then been proposed to explain such protein transition
\cite{jacob1,jacob2,garcia,onuchi,berloff,sulaiman,sulaiman2,sulaiman3}.

In our previous work, the conformational dynamics of secondary protein can be
modeled using $\phi^4$ interactions \cite{januar,januar2}. It has been shown
that the
model has reproduced the toy ad-hoc model based on the set of nonlinear
Schr\"odinger (NLS) and nonlinear Klein-Gordon equation of motions (EOMs) 
in a more natural way from first principle \cite{berloff}. In the model, the 
unfolding state of protein is initially assumed to be linear. On the other
hand, the folding pathway is induced by the nonlinear sources like
laser. Both protein conformation changes and the injected non-linear
sources are represented by the bosonic lagrangian with an additional $\phi^4$ 
interaction for the sources. It has been argued that the 'tension force' which
enables the folded pathway can be reproduced naturally in the EOM.

In this paper, we consider the nonlinear unfolding protein at the initial
state, while the external sources remain nonlinear as previously done. This is
important to investigate whether the folding mechanism and speed are influenced
by the initial conformational state. Otherwise one cannot determine if the
folded pathways are really induced and dominated by the nonlinear sources or
not.

The paper is organized as follows. First the model under consideration is
presented briefly, followed by the numerical analysis of the EOM induced by
the model. Finally it is summarized by short discussion of the results and
subsequent conclusion.

\section{The model}

This work is an extension of the previous model on protein folding
using lagrangian approach \cite{januar,januar2}. In contrast with the previous
model which assumes the initial conformational state is linear, now the protein
is initially assumed to be nonlinear likes Sine-Gordon soliton,
\be
\l_c= \frac{1}{2}\left( \pd_\mu \phi \right)^\dagger \left( \pd^\mu\phi \right)
+   \frac{m_{\phi}^4}{\lambda_{\phi}}\left[1 -
\cos\left(\frac{\sqrt{\lambda_{\phi}}}{m_{\phi}}|\phi|\right)\right] \; .
\label{eq:c}
\ee
However, the sources injected into the backbone remain nonlinear and massless.
Then, same as before the nonlinear sources are modeled by $\psi^4$ 
self-interaction.
\be
\l_s = \frac{1}{2}\left( \pd_\mu \psi \right)^\dagger  \left( \pd^\mu \psi\right) +  \frac{\lambda_{\psi}}{4!} \, (\psi^\dagger \psi)^2\;.
\label{eq:s}
\ee
The interaction term between both is described by, 
\be
\l_{int} = - \Lambda \, (\phi^\dagger \phi) (\psi^\dagger \psi)\;. 
\label{eq:int}
\ee
All of them provide the underlying model in the paper with total potential,
\be
V_{\textmd{tot}}(\psi,\phi)= \frac{m_{\phi}^4}{\lambda_{\phi}}\left[1 - \cos\left(\frac{\sqrt{\lambda_{\phi}}}{m_{\phi}}|\phi|\right)\right] + \frac{\lambda_{\psi}}{4!}\, (\psi^\dagger \psi)^2 - \Lambda \, (\phi^\dagger \phi) (\psi^\dagger \psi)\;. 
\ee

Now, throughout the paper let us assume that $\lambda_{\phi}$ is small enough,
that is approximately at the same order with $\lambda_{\psi}$. In this case, the
first term can be expanded in term of $\sqrt{\lambda_{\psi}}$,
\be
V_{\textmd{tot}}(\psi,\phi) \approx \frac{m_{\phi}^2}{2}\,\phi^\dagger \phi - \frac{\lambda_{\phi}}{4!}\,(\phi^\dagger \phi)^2 + \frac{\lambda_{\psi}}{4!}\, (\psi^\dagger \psi)^2 - \Lambda \, (\phi^\dagger \phi) (\psi^\dagger \psi)\;. 
\label{eq:Vtot}
\ee
up to the second order accuracy. If $\lambda_{\phi}=0$, the result coincides
to the case in the earlier work \cite{januar}. Imposing namely local U(1)
symmetry breaking to the total lagrangian makes the vacuum expectation value
(VEV) of the fields yields the non-trivial solutions. In the preceding model,
the 'tension force' which plays an important role to enable folded pathway has
appeared naturally by concerning the minima of total potential in term of source
field \cite{januar}. 
\be
  \langle \psi \rangle = \sqrt{\frac{12\Lambda}{\lambda_\psi}} \langle \phi
\rangle \; .
  \label{eq:vevs}
\ee
Beside of that, in term of conformation changes field the VEV is,
\be
  \langle \phi \rangle = \sqrt{\frac{6m_{\phi}^2-12\Lambda\langle \psi
\rangle^2}{\lambda_\phi}}  \; .
  \label{eq:vevc}
\ee
It shows that the existence of Sine-Gordon potential makes the early stable
ground state of conformational field turns out to be metastable. In other words,
the non trivial VEV in Eq. (\ref{eq:vevc}) constitutes new more stable ground
state of the conformational field. Transition between metastable into stable
state breaks the symmetry of the vacuum spontaneously, while the conformational
field should be nonlinear even though the external nonlinear source has not been
instilled. Therefore the protein backbone should be in nonlinear form at the
initial stage.
 
The symmetry breaking at the same time shifts the mass term of $\phi$ as follow,
\be
  m_{\phi}^2 \rightarrow \overline{m}_{\phi}^2 \equiv m_{\phi}^2 - \frac{24\Lambda^2}{\lambda_{\psi}} \langle \phi \rangle^2 \; .
\ee
Nevertheless, the nonlinear source field is set being massless, since it
represents a bunch of light source like laser.  Thus, the broken symmetry
of conformational field should not be considered to introduce its mass.

\section{The EOMs}

Having the total lagrangian at hand, one can derive the EOM using the
Euler-Lagrange equations,
\be
  \frac{\partial \l_\mathrm{tot}}{\partial |\phi|} - \partial_\mu \frac{\partial
\l_\mathrm{tot}}{\partial (|\partial_\mu \phi|)} = 0 
   \; \; \; \; \textmd{and} \; \; \; \;
\frac{\partial \l_\mathrm{tot}}{\partial |\psi|} - \partial_\mu \frac{\partial
\l_\mathrm{tot}}{\partial (|\partial_\mu \psi|)} = 0 \; ,
  \label{eq:eue}
\ee
where $\l_\mathrm{tot} = \l_c + \l_s + \l_\mathrm{int}$ in Eqs. (\ref{eq:c}),
(\ref{eq:s}) and (\ref{eq:int}) respectively.

Substituting Eqs. (\ref{eq:c}), (\ref{eq:s}) and (\ref{eq:int}) into Eq.
(\ref{eq:eue}), one immediately obtains a set of EOMs,
\bea
\frac{\partial^2|\phi|}{\partial x^2} - \frac{1}{c^2}\frac{\partial^2|\phi|}{\partial t^2} + 2\Lambda \, |\phi||\psi|^2- \frac{m_{\phi}^3c^3}{\hbar^3 \sqrt{\lambda_{\phi}}}\,\sin\left(\frac{\sqrt{\lambda_{\phi}}}{m_{\phi}}|\phi|\right)& = & 0 \; , 
\label{eq:eomc}\\
\frac{\partial^2 |\psi|}{\partial x^2} - \frac{1}{c^2}\frac{\partial^2 |\psi|}{\partial t^2} + 2 \Lambda \, |\psi||\phi|^2 - \frac{\lambda_{\psi}}{6} \, |\psi|^3 & = & 0 \; .
  \label{eq:eoms}
\eea
Here the natural unit is restored to make the light velocity $c$ and $\hbar$
reappear in the equations. The last terms in Eqs. (\ref{eq:eomc}) and
(\ref{eq:eoms}) determine the non-linearity of backbone and source 
respectively. Also, the protein mass term is melted in the Sine-Gordon
potential. One should put an attention in the second last term of Eq.
(\ref{eq:eomc}), \ie $\sim k \, \phi$ with $k \sim 2 \Lambda \langle
\psi\rangle^2$. This actually induces the tension force which is responsible
for the dynamics of conformational field and enabling the folded pathway as
expected. 

Hence, solving both EOMs in Eqs. (\ref{eq:eomc}) and (\ref{eq:eoms})
simultaneously would provide the contour of conformational changes in term of
time and one-dimensional space components. The EOMs will be solved numerically
using forward finite difference method as done in the previous work
\cite{januar,fink}. 

\section{Numerical solution of EOMs}

Same as before, it is more convenient to replace $\psi$ and $\phi$ with {\it u} and {\it w} respectively and rewritten it in explicit discrete forms as follows,
\bea
u_{i,j+1} & = & 2u_{i,j}-u_{i,j-1}+c^2\epsilon^2\left( \frac{u_{i+1,j}-2u_{i,j}+u_{i-1,j}}{\delta^2} +2\Lambda w_{i,j}^2u_{i,j}\right.\nonumber\\
&&\left.-\frac{\lambda_\psi}{6} u_{i,j}^3 \right)\;,
\label{eq:u}\\
w_{i,j+1} & = & 2w_{i,j}-w_{i,j-1}+c^2\epsilon^2 \left( \frac{w_{i+1,j}-2w_{i,j}+w_{i-1,j}}{\delta^2}+2\Lambda u_{i,j}^2w_{i,j}\right.\nonumber\\
&&\left. -\frac{m_{\phi}^3c^3}{\hbar^3\sqrt{\lambda_\phi}}\sin\left(\frac{\sqrt{\lambda_\phi}}{m_\phi} \,w_{i,j}\right) \right) \;,
\label{eq:w}
\eea
for $i = 2, 3, \cdots, N-1$ and $j = 2, 3, \cdots, M-1$.
Forward iterative procedure of the discrete EOMs can be performed if the two
lowest time values are known. First, the value at $t_1$ is fixed by the
following boundary conditions,
\be
\begin{array}{lcl}
 \psi(0,t) = \psi(L,t)=0\;\;\text{and}\;\;\phi(0,t)=\phi(L,t)=0 & \text{for} & 0\le t \le b \; , \\
\psi(x,0)=f(x)\;\;\text{and} \;\;\phi(x,0)=p(x) & \text{for} & 0\le x\le L \; , \\
\displaystyle \frac{\partial\psi(x,0)}{\partial t} = g(x)\;\;\text{and}\;\;\frac{\partial\phi(x,0)}{\partial
t}=q(x) & \text{for} & 0 < x < L \; ,
\end{array}
\label{eq:bc}
\ee
with $f(x)$, $p(x)$, $g(x)$ and $q(x)$ are newly introduced auxiliary functions. 
Secondly, the values at $\it t_2$ can be determined using second order Taylor
expansion, 
\bea
u_{i,2} & = & f_i-\epsilon g_i+\frac{c^2\epsilon^2}{2} 
\left( \frac{f_{i+1}-2f_i+f_{i-1}}{\delta^2}+2\Lambda p_i^2 f_i - \frac{\lambda_\psi}{6}
f_i^3 \right) \; , 
\label{eq:u2}\\
w_{i,2} & = & p_i-\epsilon q_i+\frac{c^2\epsilon^2}{2} 
\left( \frac{p_{i+1}-2p_i+p_{i-1}}{\delta^2}+2\Lambda
f_i^2p_i \right.\nonumber\\
&&\left. -\, \frac{m_{\phi}^3c^3}{\hbar^3\sqrt{\lambda_\phi}}\sin\left(\frac{\sqrt{\lambda_\phi}}{m_\phi} \,p_i\right)\right) \; ,
\label{eq:w2}
\eea
for $i = 2, 3, \cdots, N-1$. $\delta=\Delta x$ and $\epsilon=\Delta t$
constitutes the side length between the discretized value.

At the initial stage, suppose the nonlinear source and conformation fields have
particular form of $f(x)=2 \mathrm{sech}(2x) \, \mathrm{e}^{i2x}$ and
$g(x)=\arctan [ \exp(4x-10) ]$, while $g(x) = q(x) = 0$ for the sake
of simplicity. Furthermore, the numerical solutions can be obtained by iterative
procedure against Eqs. (\ref{eq:u}) and (\ref{eq:w}) using the results in Eqs.
(\ref{eq:u2}) and (\ref{eq:w2}) with the boundary conditions in Eq.
(\ref{eq:bc}).

\begin{figure}[t]
        \centering 
  \noindent
  \vspace*{1cm}\\
	\includegraphics[width=\textwidth]{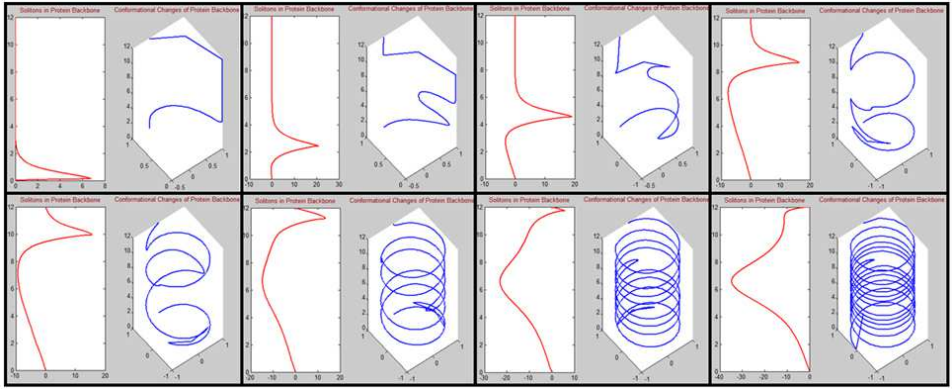}
        \caption{The soliton propagations and conformational changes on the protein backbone inducing protein folding. The vertical axis in soliton evolution denotes time in second, while the horizontal axis denotes its amplitude. The conformational changes are on the $(x,y,z)$ plane.  The constants of the simulation are chosen as $m=0.08\;eV\equiv 1.42\times 10^{-37}\;kg, L=12\;eV^{-1}\equiv 2,364\;nm,\Lambda=2.83\times 10^{-3}, \lambda_\psi=5\times 10^{-3}, \lambda_\phi=6\times 10^{-3}, \textmd{and}\; \hbar=c=1$.}
        \label{fig:1}
\end{figure}

The procedure has been performed numerically and the results are given in Fig.
\ref{fig:1}. The left figures in each box describe the propagation of
nonlinear sources in protein backbone, while the right ones show how the
protein is folded according to the time evolution. From the figure, it is clear
that the protein backbone is infinitesimally bending at the initial stage
before the nonlinear source injection. The bending constitutes the contribution
of Sine-Gordon potential into the conformation field. However, this bending is
too small to generate folding pathway, then the backbone still remains
unfolded. 

The conformation changes which generate the folding pathway start appearing
as the soliton starts propagating over the backbone. The result is
surprisingly, even slightly, different with the earlier work. The folding 
processes are slower than the linear conformation case \cite{januar}. It
might be considered as an effect of the nonlinear conformational field. One may
conclude here that the effect is destructive against the nonlinearity of 
nonlinear sources. It can also be recognized from Eq. (\ref{eq:Vtot}) that the
nonlinear terms of both fields have opposite sign.

It should be remarked that the results are obtained up to the second order
accuracy in Taylor expansion. In order to guarantee that the numerical
solutions contain no large amount of truncation errors, the step sizes
$\delta$ and $\epsilon$ are kept small enough. Nevertheless, the present method
should still be good approximation to describe visually the mechanism of
secondary protein folding. 

\section{Conclusion}

An extension of phenomenological model describing the conformational dynamics of
secondary proteins is proposed. The model is based on the matter interactions
among relevant constituents, namely the nonlinear conformational field and
the nonlinear sources. The fields are represented as the bosonic fields $\phi$
and $\psi$ in the lagrangian. It has been shown that from the bosonic lagrangian
with $\psi^4$ self-interaction, the nonlinear and tension force terms appear 
naturally as expected, and coincide with some previous works 
\cite{berloff,januar}. 

However the present model has different contour, and the folding process is
getting slower since the EOMs governing the whole dynamics
are the nonlinear Sine-Gordon and nonlinear Klein-Gordon equations. It is
argued that the nonlinearity of the both fields are against each other. Note
that the model is a generalization of earlier models which deployed both
linear, or the linear and nonlinear equations.

\section*{Acknowledgments}

MJ and AS thank the Group for Theoretical and Computational Physics LIPI for
warm hospitality during the work. This work is funded by the Indonesia Ministry
of Research and Technology and the Riset Kompetitif LIPI in fiscal year 2011
under Contract no.  11.04/SK/KPPI/II/2011.

\bibliographystyle{ws-procs975x65}
\bibliography{Januar}

\end{document}